\documentclass[11pt,a4paper,american]{extarticle}
\usepackage[T1]{fontenc}
\usepackage[latin1]{inputenc}
\usepackage{amsmath}
\usepackage{amssymb}
\usepackage{setspace}
\usepackage{esint}
\usepackage{babel}
\makeatletter

\usepackage{babel}
\usepackage{bbm}
\usepackage{hyperref}
\usepackage{authblk}
\date{}
\allowdisplaybreaks
\numberwithin{equation}{section}

\author{Renann Lipinski Jusinskas\thanks{renannlj@fzu.cz}}
\affil{Institute of Physics AS CR \authorcr  Na Slovance 2, 182 21, Praha 8\authorcr  Prague - Czech Republic}

\makeatother

\begin{document}

\title{Notes on the ambitwistor pure spinor string }
\maketitle
\begin{abstract}
In this work, some aspects of the ambitwistor pure spinor string are
investigated. The $b$ ghost is presented and its main properties
are derived in a simple way, very similar to the usual pure spinor
$b$ ghost construction. The heterotic case is also addressed with
a new proposal for the BRST charge. The BRST cohomology is shown to
correctly describe the heterotic supergravity spectrum and a semi-composite
$b$ ghost is constructed.

\tableofcontents{}
\end{abstract}

\section{Introduction\label{sec:Introduction}}

\

The so-called ambitwistor string was proposed in \cite{Mason:2013sva}
and corresponds to a chiral infinite tension limit ($\alpha'\to0$)
of the string, therefore containing only the massless spectrum. Quantization
of this model remarkably leads to the Cachazo-He-Yuan (CHY) tree level
amplitudes \cite{Cachazo:2013hca}.

Soon after Mason and Skinner's work, Berkovits came up with the pure
spinor version of the ambitwistor superstring \cite{Berkovits:2013xba},
successfully describing the CHY formulas in an explicitly supersymmetric
way, as characteristic of the pure spinor superstring.

The coupling of the RNS ambitwistor string to (NS-NS) curved backgrounds
was developed in \cite{Adamo:2014wea}, where quantum consistency
naturally imposed the non-linear $D=10$ supergravity equations of
motion.

Following an analogous idea, Chandia and Vallilo \cite{Chandia-Vallilo}
analyzed the type II supergravity background coupled to the pure spinor
string in the $\alpha'\to0$ limit and found that Berkovits' original
proposal had an extra nilpotent symmetry in the action. As it turned
out, a consistent redefinition of the pure spinor BRST charge enabled
a more natural coupling of the action to the Kalb-Ramond field and
superpartner, leading to the expected type II supergravity constraints
of \cite{Berkovits:2001ue}.

It is interesting to point out that the ambitwistor string of Mason
and Skinner have a pair of ghost fields ($b$,$\tilde{b}$) satisfying
\begin{equation}
\begin{array}{cc}
\{Q,b\}=T, & \{Q,\tilde{b}\}=H,\end{array}\label{eq:Qbbtilde}
\end{equation}
where $Q$ is the BRST charge, $T$ is the energy-momentum tensor
and $H=\frac{1}{2}P^{2}$ is the particle-like Hamiltonian. Berkovits'
pure spinor version does not seem to have a BRST-trivial energy-momentum
tensor. On the other hand, as will be shown here, the results of \cite{Chandia-Vallilo}
can be interpreted as a splitting in the holomorphic theory which
is responsible for a very simple construction of the $b$ ghost and
a generalization of the operator $\tilde{b}$ of \eqref{eq:Qbbtilde}.
In simple terms, one can define for each ``sector'' a new field,
$b_{+}$ and $b_{-}$, satisfying $\{Q,b_{\pm}\}=T_{\pm}$, with\begin{subequations}
\begin{eqnarray}
T_{+}+T_{-} & = & T,\\
T_{+}-T_{-} & = & \frac{1}{2}P^{2}+\ldots.\label{eq:T-T}
\end{eqnarray}
\end{subequations}The dots in \eqref{eq:T-T} are extra terms required
to make the right hand side BRST-closed. The operators $b_{+}$ and
$b_{-}$ are very similar to the composite pure spinor $b$ ghost
but their geometrical interpretation is not clear yet. Unlike in Berkovits'
proposal, a concrete form for the integrated vertex operator is still
lacking in Chandia and Vallilo's modification and a better understanding
on the newly introduced $b_{+}$ and $b_{-}$ might help to solve
this issue.

Concerning the heterotic case, also proposed in \cite{Berkovits:2013xba},
the energy-momentum tensor is clearly BRST-trivial but there does
not seem to exist a $\tilde{b}$ operator trivializing the particle-like
Hamiltonian. Maybe a bit more worrying is the fact that the supergravity
states do not have a satisfactory vertex operator description.

Motivated by the holomorphic sectorization of the type II case, the
heterotic BRST charge will be modified to
\begin{equation}
Q=\oint\,\{\lambda^{\alpha}d_{\alpha}+\overline{c}T_{+}-\overline{b}\overline{c}\partial\overline{c}\},\label{eq:newhetQ}
\end{equation}
where $\lambda^{\alpha}$ is the pure spinor ghost, $d_{\alpha}$
is the improved operator proposed in \cite{Chandia-Vallilo}, ($\overline{b}$,$\overline{c}$)
are the reparametrization ghosts and $T_{+}$ is a \emph{fake} energy-momentum
satisfying\begin{subequations}
\begin{eqnarray}
T_{+}(x)\,T_{+}(y) & \sim & \frac{2T_{+}}{(z-y)^{2}}+\frac{\partial T_{+}}{(z-y)},\\
T_{+}(x)\,\lambda^{\alpha}d_{\alpha}(y) & \sim & \textrm{regular}.
\end{eqnarray}
\end{subequations}Besides having the ($b$,$\tilde{b}$) structure
mentioned above, the BRST charge of \eqref{eq:newhetQ} will be shown
to correctly describe the massless heterotic spectrum (super Yang-Mills
and supergravity). In terms of the redefined supersymmetric invariants,
the heterotic action will be rewritten such that the coupling with
the Kalb-Ramond field is manifest, exactly like in the type II case.

This work is organized as follows. Section \ref{sec:typeII} discusses
the type II case of the infinite tension limit of the pure spinor
string. After a review of Berkovits' proposal and the modification
proposed by Chandia and Vallilo, the holomorphic sectorization is
studied and the construction of the composite $b$ ghost is presented
in detail, together with several properties. Section \ref{sec:heterotic}
starts with a review of the heterotic case, explaining why the natural
choice for the supergravity vertex is incomplete. With the new proposal
for the BRST charge, this flaw is corrected and a semi-composite $b$
ghost is introduced. Section \ref{sec:remarks} discusses the results
and possible directions to follow.

\section{The type II ambitwistor pure spinor string\label{sec:typeII}}

\

The $\alpha'\to0$ limit of the pure spinor superstring was first
discussed in \cite{Berkovits:2013xba}. For the type II case, the
proposed action is simply
\begin{equation}
S=\int d^{2}z\{P_{m}\overline{\partial}X^{m}+p_{\alpha}\overline{\partial}\theta^{\alpha}+w_{\alpha}\overline{\partial}\lambda^{\alpha}+\hat{p}_{\hat{\alpha}}\overline{\partial}\hat{\theta}^{\hat{\alpha}}+\hat{w}_{\hat{\alpha}}\overline{\partial}\hat{\lambda}^{\hat{\alpha}}\},\label{eq:typeIIaction}
\end{equation}
where \{$P_{m}$, $p_{\alpha}$, $\hat{p}_{\hat{\alpha}}$\} denote
the conjugate momenta to the $\mathcal{N}=2$ superspace coordinates
\{$X^{m}$, $\theta^{\alpha}$, $\hat{\theta}^{\hat{\alpha}}$\},
and $(w_{\alpha},\lambda^{\alpha})$ and $(\hat{w}_{\hat{\alpha}},\hat{\lambda}^{\hat{\alpha}})$
are the usual pure spinor ghost conjugate pairs. For convenience,
the same chirality is being considered for the superspace coordinates
$\theta$ and $\hat{\theta}$ (type IIB) but the results are easily
generalized to the type IIA case.

The first order action $S$ is supersymmetric with respect to the
charges\begin{subequations}
\begin{eqnarray}
q_{\alpha} & = & \oint\{p_{\alpha}+\frac{1}{2}P_{m}(\gamma^{m}\theta)_{\alpha}\},\\
\hat{q}_{\hat{\alpha}} & = & \oint\{\hat{p}_{\hat{\alpha}}+\frac{1}{2}P_{m}(\gamma^{m}\hat{\theta})_{\hat{\alpha}}\},
\end{eqnarray}
\end{subequations}which define the invariants $P_{m}$ and\begin{subequations}
\begin{eqnarray}
\Pi^{m} & = & \partial X^{m}+\frac{1}{2}(\theta\gamma^{m}\partial\theta)+\frac{1}{2}(\hat{\theta}\gamma^{m}\partial\hat{\theta}),\label{eq:Pisusyinv}\\
d_{\alpha} & = & p_{\alpha}-\frac{1}{2}P_{m}(\gamma^{m}\theta)_{\alpha},\label{eq:dsusyinv}\\
\hat{d}_{\hat{\alpha}} & = & \hat{p}_{\hat{\alpha}}-\frac{1}{2}P_{m}(\gamma^{m}\hat{\theta})_{\hat{\alpha}}.
\end{eqnarray}
\end{subequations}

As usual, $S$ has to be provided with the BRST charge
\begin{equation}
Q=\oint\{\lambda^{\alpha}d_{\alpha}+\hat{\lambda}^{\hat{\alpha}}\hat{d}_{\hat{\alpha}}\},\label{eq:BRSTnb}
\end{equation}
Nilpotency of $Q$ follows from the pure spinor constraints $(\lambda\gamma^{m}\lambda)=(\hat{\lambda}\gamma^{m}\hat{\lambda})=0$. 

As expected, the type II supergravity spectrum is in the cohomology
of \eqref{eq:BRSTnb}. BRST-closedness of the vertex
\begin{equation}
U_{SG}=\lambda^{\alpha}\hat{\lambda}^{\hat{\alpha}}A_{\alpha\hat{\alpha}}(\theta,\hat{\theta})e^{ik_{m}X^{m}},\label{eq:sugraIIvertex}
\end{equation}
imply the linearized supergravity equations of motion for the superfield
$A_{\alpha\hat{\alpha}}$:
\begin{equation}
\begin{array}{ccc}
\gamma_{mnpqr}^{\alpha\beta}D_{\beta}A_{\alpha\hat{\alpha}}=0, & \; & \gamma_{mnpqr}^{\hat{\alpha}\hat{\beta}}\hat{D}_{\hat{\beta}}A_{\alpha\hat{\alpha}}=0.\end{array}\label{eq:SGeomTII}
\end{equation}
Here, $D_{\alpha}\equiv\partial_{\alpha}+\frac{i}{2}(\gamma^{m}\theta)_{\alpha}k_{m}$
and $\hat{D}_{\hat{\alpha}}=\partial_{\hat{\alpha}}+\frac{i}{2}(\gamma^{m}\hat{\theta})_{\hat{\alpha}}k_{m}$
are the supersymmetric derivatives for momentum eigenstates. The gauge
transformations come from the BRST-exact states of the form $\Lambda=\lambda^{\alpha}\Lambda_{\alpha}+\hat{\lambda}^{\hat{\alpha}}\hat{\Lambda}_{\hat{\alpha}}$,
implying the gauge transformation $\delta A_{\alpha\hat{\alpha}}=D_{\alpha}\hat{\Lambda}_{\hat{\alpha}}+\hat{D}_{\hat{\alpha}}\Lambda_{\alpha}$,
as long as the superfield parameters satisfy $D\gamma^{mnpqr}\Lambda=\hat{D}\gamma^{mnpqr}\hat{\Lambda}=0$. 

For convenience, $A_{\alpha\hat{\alpha}}$ in \eqref{eq:sugraIIvertex}
can be cast as $A_{\alpha\hat{\alpha}}=A_{\alpha}(\theta)\hat{A}_{\hat{\alpha}}(\hat{\theta})$,
such that one can introduce the usual auxiliary fields satisfying\begin{subequations}
\begin{eqnarray}
A_{m} & = & \frac{1}{8}(D_{\alpha}\gamma_{m}^{\alpha\beta}A_{\alpha}),\\
W^{\alpha} & = & \frac{1}{10}\gamma_{m}^{\alpha\beta}[D_{\beta}A_{m}-ik^{m}\gamma_{m}^{\alpha\beta}A_{\beta}],\\
F_{mn} & = & \frac{i}{2}\left(k_{m}A_{n}-k_{n}A_{m}\right),
\end{eqnarray}
\end{subequations}and similar equations for $\hat{A}_{m}$, $\hat{W}^{\hat{\alpha}}$
and $\hat{F}_{mn}$ in terms of $\hat{A}_{\hat{\alpha}}$. These auxiliary
fields are the basic ingredients of the integrated vertex $V$ presented
in \cite{Berkovits:2013xba}, given by:
\begin{equation}
V=\int d^{2}z\,\overline{\delta}(k^{m}P_{m})[P_{m}A^{m}+d_{\alpha}W^{\alpha}+N^{mn}F_{mn}][P_{m}\hat{A}^{m}+\hat{d}_{\hat{\alpha}}\hat{W}^{\hat{\alpha}}+\hat{N}^{mn}\hat{F}_{mn}]e^{ik_{m}X^{m}}.\label{eq:integratedvertexB}
\end{equation}
$N_{mn}$ and $\hat{N}_{mn}$ are the ghost Lorentz currents, defined
as
\begin{equation}
\begin{array}{ccc}
N^{mn}\equiv\frac{1}{2}(\lambda\gamma^{mn}w), & \; & \hat{N}^{mn}\equiv\frac{1}{2}(\hat{\lambda}\gamma^{mn}\hat{w}),\end{array}\label{eq:Lorentzghost}
\end{equation}
and the operator $\overline{\delta}(k^{m}P_{m})$ is detailedly described
in \cite{Mason:2013sva}, having the right conformal dimensions necessary
to make $V$ a worldsheet scalar. Observe that BRST-closedness and
gauge transformations of $V$ ($\delta A_{m}=k_{m}\Lambda$ and $\delta\hat{A}_{m}=k_{m}\hat{\Lambda}$)
can be shown to be proportional to $\overline{\delta}(k\cdot P)\,k^{m}P_{m}$.

The pure spinor tree level amplitudes computed using the massless
vertices described above have explicit spacetime supersymmetry and
were shown to agree with the RNS computations \cite{Berkovits:2013xba,Gomez:2013wza}. 

In spite of the interesting outcomes, Berkovits' proposal has yet
to be better understood. The BRST cohomology of \eqref{eq:BRSTnb}
is not clear enough and a consistent coupling with curved backgrounds
seems to require a slight modification of the flat space limit just
presented \cite{Chandia-Vallilo}. These features will be discussed,
reviewed and extended in the rest of the section.

\subsection{Extra elements in the BRST cohomology\label{sub:extrastates}}

\

The simple form of the BRST charge \ref{eq:BRSTnb} hides a fundamental
feature of the closed string spectrum that is the decoupling of the
left-moving and right-moving sectors. Of course, the \emph{chiral}
action \eqref{eq:typeIIaction} is not able to encode this information
and this has an interesting consequence, as there might be extra states
in the BRST cohomology.

Most of the cohomology analysis for the $\alpha'\to0$ limit reviewed
above can be parallelized with the $\mathcal{N}=2$ pure spinor superparticle.
In \cite{Berkovits:2002uc} there is a thorough discussion on the
physical spectrum coming from the quantization of the superparticle,
in particular that of the zero-momentum states. Of course, to talk
about \emph{physical} spectrum one has to define the \emph{physical}
state conditions. This will be discussed in section \ref{sec:remarks}
because it is fundamentally related to the developments to be presented
in the next subsections.

For now, it will be pointed out that at zero-momentum there are also
non-vanishing conformal weight states in the BRST cohomology. Consider,
for example, the operator
\[
(\lambda\gamma^{m}\partial\theta),
\]
which is BRST-closed and have conformal weight one. In the full superstring
(finite $\alpha'$), it would correspond to the BRST transformation
of the operator $\Pi^{m}$. However here,
\begin{equation}
[Q,\Pi^{m}]=(\lambda\gamma^{m}\partial\theta)+(\hat{\lambda}\gamma^{m}\partial\hat{\theta}).
\end{equation}
In fact, there does not seem to exist an operator $O^{m}$ such that
$[Q,O^{m}]=(\lambda\gamma^{m}\partial\theta)$. The same holds for
$(\hat{\lambda}\gamma^{m}\partial\hat{\theta})$.

Usually, BRST-closed states with nonvanishing conformal weight can
be argued to be BRST-exact. This follows from the fact that the energy-momentum
tensor itself is BRST-trivial, \emph{i.e.} this argument relies on
the existence of a $b$ ghost satisfying $\{Q,b\}=T$. For the action
\eqref{eq:typeIIaction}, the energy momentum-tensor is given by
\begin{equation}
T=-P_{m}\partial X^{m}-p_{\alpha}\partial\theta^{\alpha}-\hat{p}_{\hat{\alpha}}\partial\hat{\theta}^{\hat{\alpha}}-w_{\alpha}\partial\lambda^{\alpha}-\hat{w}_{\hat{\alpha}}\partial\hat{\lambda}^{\hat{\alpha}},\label{eq:energy-momentum}
\end{equation}
and the known procedure to build the composite pure spinor $b$ ghost
\cite{Berkovits:2004px,Berkovits:2005bt} does not work here. This
will be clarified soon, but technically it is related to the mixing
of the variables that would describe the left and right-moving sector
of the finite tension superstring. 

The above observation raises the question about massive states, which
are usually built out of non-vanishing conformal weight fields. Since
the operators of the form $\exp(ik_{m}X^{m})$ are worldsheet scalars
in the $\alpha'\to0$ limit, in a BRST trivial energy-momentum scenario
this would mean that the cohomology consists of massless states only
. On the other hand, the action $S$ has further symmetries. One of
them, in particular, is generated by the particle-like Hamiltonian
\begin{equation}
\mathcal{H}_{B}=-\frac{1}{2}P_{m}P^{m},\label{eq:HB}
\end{equation}
which can be interpreted as the mass operator and commutes with the
BRST charge. If one requires the physical states to be annihilated
by $\mathcal{H}_{B}$, that would automatically project out any possible
massive BRST-closed state.

As it turns out, $\mathcal{H}_{B}$ is BRST-exact \cite{Berkovits:2015yra}.
To show that, consider first the following:
\begin{equation}
\begin{array}{cc}
g^{\alpha}\equiv\frac{1}{4}(\gamma^{m}d)^{\alpha}P_{m}, & \{Q,g^{\alpha}\}=\frac{1}{2}\lambda^{\alpha}\mathcal{H}_{B},\\
\hat{g}^{\hat{\alpha}}\equiv\frac{1}{4}(\gamma^{m}\hat{d})^{\hat{\alpha}}P_{m}, & \{Q,\hat{g}^{\hat{\alpha}}\}=\frac{1}{2}\hat{\lambda}^{\hat{\alpha}}\mathcal{H}_{B}.
\end{array}
\end{equation}
Next, defining\begin{subequations}
\begin{eqnarray}
B^{+} & \equiv & \frac{C\cdot g}{C\cdot\lambda}+\frac{\hat{C}\cdot\hat{g}}{\hat{C}\cdot\hat{\lambda}},\\
B^{-} & \equiv & \frac{C\cdot g}{C\cdot\lambda}-\frac{\hat{C}\cdot\hat{g}}{\hat{C}\cdot\hat{\lambda}},
\end{eqnarray}
\end{subequations}for any nonvanishing $(C\cdot\lambda)$ and $(\hat{C}\cdot\hat{\lambda})$,
with $C_{\alpha}$ and $\hat{C}_{\hat{\alpha}}$ constant spinors,
it can be demonstrated that\begin{subequations}\label{eq:B+B-}
\begin{eqnarray}
\{Q,B^{+}\} & = & \mathcal{H}_{B},\\
\{Q,B^{-}\} & = & 0.
\end{eqnarray}
\end{subequations}In particular, it implies that any BRST-closed
eigenstate of $\mathcal{H}_{B}$ with nonzero eigenvalue is BRST-exact.
The operator $B^{-}$ is BRST-closed and the absence of a $b$ ghost
makes it hard to tell whether it is BRST-exact, although unlikely.
The covariant versions of these operators would require the introduction
of the nonminimal sector \cite{Berkovits:2005bt} and have been defined
also in \cite{Adamo:2015hoa}, similarly to what is done in subsection
\ref{sub:typeIIbghost}.

Altogether, these observations indicate that the original proposal
of \cite{Berkovits:2013xba} might be incomplete, since the BRST cohomology
is enhanced when compared to the zero-momentum spectrum of the superstring
and it is not clear whether this is relevant for a well-defined worldsheet
theory for supergravity.

In fact, Chandia and Vallilo \cite{Chandia-Vallilo} already considered
this possibility from another perspective. In an attempt to obtain
the supergravity constraints from a consistent coupling of the type
II background to the free action \eqref{eq:typeIIaction}, they noticed
another symmetry which led to a modification of the flat space limit
and a redefinition of the supersymmetry charge and consequently the
operators $d_{\alpha}$ and $\hat{d}_{\hat{\alpha}}$. This will be
reviewed next.

\subsection{Review of the improved BRST-charge\label{sub:reviewCV}}

\

The key observation in \cite{Chandia-Vallilo} is that the action
$S$ is also invariant under another nilpotent symmetry generated
by
\begin{equation}
\mathcal{K}\equiv\oint\,\{(\lambda\gamma_{m}\theta)[\partial X^{m}+\frac{1}{2}(\theta\gamma^{m}\partial\theta)]-(\hat{\lambda}\gamma_{m}\hat{\theta})[\partial X^{m}+\frac{1}{2}(\hat{\theta}\gamma^{m}\partial\hat{\theta})]\}.
\end{equation}
Although the two terms (hatted and unhatted) above are independent
symmetries of the action, only this particular combination is BRST-closed.
Concerning supersymmetry, it is easy to show that $\mathcal{K}$ is
supersymmetric up to BRST-exact terms:\begin{subequations}
\begin{eqnarray}
\{q_{\alpha},\mathcal{K}\} & = & \{Q,\oint\,(\gamma_{m}\theta)_{\alpha}[\partial X^{m}+\frac{1}{2}(\theta\gamma^{m}\partial\theta)]\},\\
\{\hat{q}_{\hat{\alpha}},\mathcal{K}\} & = & -\{Q,\oint\,(\gamma_{m}\hat{\theta})_{\hat{\alpha}}[\partial X^{m}+\frac{1}{2}(\hat{\theta}\gamma^{m}\partial\hat{\theta})]\}.
\end{eqnarray}
\end{subequations}

Based on Berkovits' suggestion that $Q+\mathcal{K}$ should be the
BRST charge instead, Chandia and Vallilo made a consistent redefinition
of the supersymmetry charges and supersymmetric invariants\footnote{In fact, the action \eqref{eq:typeIIaction} has two other nilpotent
symmetries, generated by $\mathcal{K}_{1}=\oint\,(\lambda\gamma_{m}\theta)(\hat{\theta}\gamma^{m}\partial\hat{\theta})$
and $\mathcal{K}_{2}=\oint\,(\hat{\lambda}\gamma_{m}\hat{\theta})(\theta\gamma^{m}\partial\theta)$,
but there does not seem to be any operator redefinition consistent
with $\mathcal{N}=2$ supersymmetry that would incorporate them, as
they mix the spinor chiralities.}. The operators $d_{\alpha}$ and $\hat{d}_{\hat{\alpha}}$ were redefined
as\begin{subequations}\label{eq:newdCV}
\begin{eqnarray}
d_{\alpha} & \equiv & p_{\alpha}-\frac{1}{2}(P_{m}-\partial X_{m})(\gamma^{m}\theta)_{\alpha}+\frac{1}{4}(\theta\gamma_{m}\partial\theta)(\gamma^{m}\theta)_{\alpha},\\
\hat{d}_{\hat{\alpha}} & \equiv & \hat{p}_{\hat{\alpha}}-\frac{1}{2}(P_{m}+\partial X_{m})(\gamma^{m}\hat{\theta})_{\hat{\alpha}}-\frac{1}{4}(\hat{\theta}\gamma_{m}\partial\hat{\theta})(\gamma^{m}\hat{\theta})_{\hat{\alpha}},
\end{eqnarray}
\end{subequations}together with the supersymmetry charges\begin{subequations}\label{eq:newsusyCV}
\begin{eqnarray}
q_{\alpha} & \equiv & \oint\{p_{\alpha}+\frac{1}{2}(P_{m}-\partial X_{m})(\gamma^{m}\theta)_{\alpha}-\frac{1}{12}(\theta\gamma_{m}\partial\theta)(\gamma^{m}\theta)_{\alpha}\},\\
\hat{q}_{\hat{\alpha}} & \equiv & \oint\{\hat{p}_{\hat{\alpha}}+\frac{1}{2}(P_{m}+\partial X_{m})(\gamma^{m}\hat{\theta})_{\hat{\alpha}}+\frac{1}{12}(\hat{\theta}\gamma_{m}\partial\hat{\theta})(\gamma^{m}\hat{\theta})_{\hat{\alpha}}\}.
\end{eqnarray}
\end{subequations} 

$P_{m}$ is no longer invariant under supersymmetry, only the combination
\[
P_{m}-\frac{1}{2}(\theta\gamma_{m}\partial\theta)+\frac{1}{2}(\hat{\theta}\gamma_{m}\partial\hat{\theta}).
\]
It is convenient, however, to write it in a linear combination with
$\Pi^{m}$ defined in \eqref{eq:Pisusyinv}, introducing two other
supersymmetric invariants that will appear naturally in the OPE algebra:\begin{subequations}\label{eq:newPCV}
\begin{eqnarray}
P_{m}^{-} & \equiv & P_{m}-\partial X_{m}-(\theta\gamma_{m}\partial\theta),\\
P_{m}^{+} & \equiv & P_{m}+\partial X_{m}+(\hat{\theta}\gamma_{m}\partial\hat{\theta}).
\end{eqnarray}
\end{subequations}

The action $S$ in \eqref{eq:typeIIaction} can be rewritten in terms
of the newly defined operators as
\begin{eqnarray}
S & = & \int d^{2}z\{\frac{1}{2}(P_{m}^{+}+P_{m}^{-})\overline{\Pi}^{m}+d_{\alpha}\overline{\partial}\theta^{\alpha}+w_{\alpha}\overline{\partial}\lambda^{\alpha}+\hat{d}_{\hat{\alpha}}\overline{\partial}\hat{\theta}^{\hat{\alpha}}+\hat{w}_{\hat{\alpha}}\overline{\partial}\hat{\lambda}^{\hat{\alpha}}\}\nonumber \\
 &  & -\frac{1}{2}\int d^{2}z\{\Pi_{m}[(\theta\gamma^{m}\overline{\partial}\theta)-(\hat{\theta}\gamma^{m}\overline{\partial}\hat{\theta})]-[(\theta\gamma^{m}\partial\theta)-(\hat{\theta}\gamma^{m}\partial\hat{\theta})]\overline{\Pi}_{m}\}\nonumber \\
 &  & -\frac{1}{4}\int d^{2}z\{(\theta\gamma_{m}\partial\theta)(\hat{\theta}\gamma^{m}\overline{\partial}\hat{\theta})-(\hat{\theta}\gamma_{m}\partial\hat{\theta})(\theta\gamma^{m}\overline{\partial}\theta)\},
\end{eqnarray}
where $\overline{\Pi}^{m}$ is just the antiholomorphic version of
$\Pi^{m}$. The BRST charge $Q$ has the same form \eqref{eq:BRSTnb},
but now with the modified $d_{\alpha}$ and $\hat{d}_{\hat{\alpha}}$
of \eqref{eq:newdCV}. It is worth to point out the the integrated
vertex displayed in \eqref{eq:integratedvertexB} is no longer BRST-closed
with respect to the modified charge and this is so far an unsolved
issue.

The relevant OPE's for the improved set of operators can be summarized
as
\begin{equation}
\begin{array}{rclcrcl}
d_{\alpha}(z)d_{\beta}(y) & \sim & -\frac{P_{m}^{-}\gamma_{\alpha\beta}^{m}}{(z-y)}, &  & \hat{d}_{\hat{\alpha}}(z)\hat{d}_{\hat{\beta}}(y) & \sim & -\frac{P_{m}^{+}\gamma_{\hat{\alpha}\hat{\beta}}^{m}}{(z-y)},\\
 &  &  & \qquad\\
d_{\alpha}(z)P_{m}^{-}(y) & \sim & -2\frac{(\gamma_{m}\partial\theta)_{\alpha}}{(z-y)}, &  & \hat{d}_{\hat{\alpha}}(z)P_{m}^{+}(y) & \sim & 2\frac{(\gamma_{m}\partial\hat{\theta})_{\hat{\alpha}}}{(z-y)},\\
\\
P_{m}^{-}(z)P_{n}^{-}(y) & \sim & 2\frac{\eta_{mn}}{(z-y)^{2}}, &  & P_{m}^{+}(z)P_{n}^{+}(y) & \sim & -2\frac{\eta_{mn}}{(z-y)^{2}},\\
\\
d_{\alpha}(z)\Pi^{m}(y) & \sim & \frac{(\gamma_{m}\partial\theta)_{\alpha}}{(z-y)}, &  & \hat{d}_{\hat{\alpha}}(z)\Pi^{m}(y) & \sim & \frac{(\gamma_{m}\partial\hat{\theta})_{\hat{\alpha}}}{(z-y)},\\
\\
P_{m}^{-}(z)\Pi^{n}(y) & \sim & -\frac{\delta_{m}^{n}}{(z-y)^{2}}, &  & P_{m}^{+}(z)\Pi^{n}(y) & \sim & -\frac{\delta_{m}^{n}}{(z-y)^{2}}.
\end{array}
\end{equation}

Notice that there is a clear splitting and the two sectors $\{P_{m}-\partial X_{m},p_{\alpha},\theta^{\alpha},w_{\alpha},\lambda^{\alpha}\}$
and $\{P_{m}+\partial X_{m},\hat{p}_{\hat{\alpha}},\hat{\theta}^{\hat{\alpha}},\hat{w}_{\hat{\alpha}},\hat{\lambda}^{\hat{\alpha}}\}$
are ``decoupled''. Next subsection will extend this idea and introduce
the pure spinor $b$ ghost for the type II ambitwistor string.

\subsection{Holomorphic sectorization and the $b$ ghost\label{sub:typeIIbghost}}

\

The proposal of \cite{Chandia-Vallilo} splits the chiral action $S$
in two sectors which emulate the would-be left and right-moving sectors
of the superstring. It can be shown that this feature easily solves
the cohomology issues discussed in subsection \ref{sub:extrastates}.
In fact it enables a very simple construction for the composite $b$
ghost. To do that, the two sectors have to be better understood.

It is interesting to observe, for example, that the energy-momentum
tensor of \eqref{eq:energy-momentum} can be written in a way that
makes this splitting explicit. Using the operators defined in \eqref{eq:newdCV}
and \eqref{eq:newPCV}, $T$ is written as
\begin{equation}
T=T_{+}+T_{-},\label{eq:emT}
\end{equation}
where\begin{subequations}
\begin{eqnarray}
T_{-} & \equiv & \frac{1}{4}\eta^{mn}P_{m}^{-}P_{n}^{-}-d_{\alpha}\partial\theta^{\alpha}-w_{\alpha}\partial\lambda^{\alpha},\\
T_{+} & \equiv & -\frac{1}{4}\eta^{mn}P_{m}^{+}P_{n}^{+}-\hat{d}_{\hat{\alpha}}\partial\hat{\theta}^{\hat{\alpha}}-\hat{w}_{\hat{\alpha}}\partial\hat{\lambda}^{\hat{\alpha}}.
\end{eqnarray}
\end{subequations}Both $T_{-}$ and $T_{+}$ are BRST-closed and
can be viewed as \emph{fake} anomaly-free energy-momentum tensors
for each sector\emph{}\footnote{One has to be careful with this interpretation because only the linear
combination in \eqref{eq:emT} has the expected properties of a energy-momentum
tensor when acting on $X^{m}$ or $P_{m}$, \emph{e.g.
\[
\begin{array}{ccc}
T_{+}(x)\,X^{m}(y)\sim\frac{1}{2}\frac{(\partial X^{m}+P^{m})}{(z-y)}, & \; & T_{-}(x)\,X^{m}(y)\sim\frac{1}{2}\frac{(\partial X^{m}-P^{m})}{(z-y)}.\end{array}
\]
}}:\begin{subequations}\label{eq:partialTOPES}
\begin{eqnarray}
T_{-}(x)T_{-}(y) & \sim & \frac{2T_{-}}{(z-y)^{2}}+\frac{\partial T_{-}}{(z-y)},\\
T_{+}(x)T_{+}(y) & \sim & \frac{2T_{+}}{(z-y)^{2}}+\frac{\partial T_{+}}{(z-y)},\\
T_{-}(x)T_{+}(y) & \sim & \textrm{regular}.
\end{eqnarray}
\end{subequations}

Note that $\mathcal{H}_{B}$ is not BRST-closed with respect to the
new BRST-charge, which comes from the fact that $[\mathcal{K},\mathcal{H}_{B}]\neq0$
in subsection \ref{sub:reviewCV}. However, one can define
\begin{equation}
\mathcal{H}_{CV}\equiv T_{+}-T_{-},\label{eq:Hcv}
\end{equation}
which is interpreted as a generalization of $\mathcal{H}_{B}$ in
\eqref{eq:HB} \cite{Chandia-Vallilo}. A natural question is whether
$\mathcal{H}_{CV}$ is BRST-exact or not. If so, given the sectorization
so far observed, it is likely that both $T_{+}$ and $T_{-}$ are
BRST-exact, leading to a trivialization of the energy-momentum tensor.

Motivated by the original proposal for the pure spinor $b$ ghost
\cite{Berkovits:2004px}, one can define the operators\begin{subequations}\label{eq:Gdef}
\begin{eqnarray}
G^{\alpha} & \equiv & -\frac{1}{4}\eta^{mn}\gamma_{m}^{\alpha\beta}(d_{\beta},P_{m}^{-})-\frac{1}{4}N_{mn}(\gamma^{mn}\partial\theta)^{\alpha}-\frac{1}{4}J\partial\theta^{\alpha}-\partial^{2}\theta^{\alpha},\\
\hat{G}^{\hat{\alpha}} & \equiv & \frac{1}{4}\eta^{mn}\gamma_{m}^{\hat{\alpha}\hat{\beta}}(\hat{d}_{\hat{\beta}},P_{m}^{+})-\frac{1}{4}\hat{N}_{mn}(\gamma^{mn}\partial\hat{\theta})^{\hat{\alpha}}-\frac{1}{4}\hat{J}\partial\hat{\theta}^{\hat{\alpha}}-\partial^{2}\hat{\theta}^{\hat{\alpha}}.
\end{eqnarray}
\end{subequations}$N_{mn}$ and $\hat{N}_{mn}$ are ghost Lorentz
currents displayed in \eqref{eq:Lorentzghost}, and $J$ and $\hat{J}$
are the ghost number currents:
\begin{equation}
\begin{array}{ccc}
J\equiv-w\cdot\lambda, & \; & \hat{J}\equiv-\hat{w}\cdot\hat{\lambda}.\end{array}\label{eq:ghostnumber}
\end{equation}
Observe that one has to take into account quantum effects of non-commuting
operators and the ordering prescription that will be adopted here
is
\begin{equation}
(A,B)(y)\equiv\frac{1}{2\pi i}\oint\frac{dz}{z-y}A(z)B(y).\label{eq:ordering}
\end{equation}
It is straightforward to show that the operators in \eqref{eq:Gdef}
satisfy the following properties,\begin{subequations}
\begin{eqnarray*}
\{Q,G^{\alpha}\} & = & (\lambda^{\alpha},T_{-}),\\
\{Q,\hat{G}^{\hat{\alpha}}\} & = & (\hat{\lambda}^{\hat{\alpha}},T_{+}),
\end{eqnarray*}
\end{subequations}resembling the usual holomorphic construction.

In order to present a covariant version of the $b$ ghost, the known
chain of operators introduced in \cite{Berkovits:2004px,Berkovits:2005bt}
will be mirrored here. In fact there is little to change, only some
overall factors. These operators are defined as\begin{subequations}\label{eq:HKLdef}
\begin{eqnarray}
H^{\alpha\beta} & \equiv & -\frac{1}{768}\gamma_{mnp}^{\alpha\beta}(d\gamma^{mnp}d+24N^{mn}\eta^{pq}P_{q}^{-}),\\
\hat{H}^{\hat{\alpha}\hat{\beta}} & \equiv & \frac{1}{768}\gamma_{mnp}^{\hat{\alpha}\hat{\beta}}(\hat{d}\gamma^{mnp}\hat{d}+24\hat{N}^{mn}\eta^{pq}P_{q}^{+}),\\
K^{\alpha\beta\gamma} & \equiv & \frac{1}{192}N_{mn}\gamma_{mnp}^{[\alpha\beta}(\gamma^{p}d)^{\gamma]},\\
\hat{K}^{\hat{\alpha}\hat{\beta}\hat{\gamma}} & \equiv & -\frac{1}{192}\hat{N}_{mn}\gamma_{mnp}^{[\hat{\alpha}\hat{\beta}}(\gamma^{p}\hat{d})^{\hat{\gamma}]},\\
L^{\alpha\beta\gamma\lambda} & \equiv & \frac{1}{6144}(N^{mn},N^{rs})\eta^{pq}\gamma_{mnp}^{[\alpha\beta}\gamma_{qrs}^{\gamma]\lambda},\\
\hat{L}^{\hat{\alpha}\hat{\beta}\hat{\gamma}\hat{\lambda}} & \equiv & -\frac{1}{6144}(\hat{N}^{mn},\hat{N}^{rs})\eta^{pq}\gamma_{mnp}^{[\hat{\alpha}\hat{\beta}}\gamma_{qrs}^{\hat{\gamma}]\hat{\lambda}},
\end{eqnarray}
\end{subequations}and satisfy
\begin{equation}
\begin{array}{rclrcl}
[Q,H^{\alpha\beta}] & = & (\lambda^{[\alpha},G^{\beta]}), & [Q,\hat{H}^{\hat{\alpha}\hat{\beta}}] & = & (\hat{\lambda}^{[\hat{\alpha}},G^{\hat{\beta}]}),\\
\{Q,K^{\alpha\beta\gamma}\} & = & (\lambda^{[\alpha},H^{\beta\gamma]}), & \{Q,\hat{K}^{\hat{\alpha}\hat{\beta}\hat{\gamma}}\} & = & (\hat{\lambda}^{[\hat{\alpha}},\hat{H}^{\hat{\beta}\hat{\gamma}]}),\\{}
[Q,L^{\alpha\beta\gamma\lambda}] & = & (\lambda^{[\alpha},K^{\beta\gamma\lambda]}), & [Q,\hat{L}^{\hat{\alpha}\hat{\beta}\hat{\gamma}\hat{\lambda}}] & = & (\hat{\lambda}^{[\hat{\alpha}},\hat{K}^{\hat{\beta}\hat{\gamma}\hat{\lambda}]}),\\
(\lambda^{[\alpha},L^{\beta\gamma\lambda\sigma]}) & = & 0, & (\hat{\lambda}^{[\hat{\alpha}},\hat{L}^{\hat{\beta}\hat{\gamma}\hat{\lambda}\hat{\sigma}]}) & = & 0.
\end{array}\label{eq:QHKL}
\end{equation}
The square brackets denote indices antisymmetrization and it can be
read as
\begin{equation}
\left[\alpha_{1}\ldots\alpha_{n}\right]=\frac{1}{n!}\left(\alpha_{1}\ldots\alpha_{n}+\textrm{all antisymmetric permutations}\right).
\end{equation}

The next step is to introduce the non-minimal variables of \cite{Berkovits:2005bt},
which enter the action as
\begin{equation}
S_{nm}=\int d^{2}z\{\overline{w}^{\alpha}\overline{\partial}\overline{\lambda}_{\alpha}+s^{\alpha}\overline{\partial}r_{\alpha}+\hat{\overline{w}}^{\hat{\alpha}}\overline{\partial}\hat{\overline{\lambda}}_{\hat{\alpha}}+\hat{s}^{\hat{\alpha}}\overline{\partial}\hat{r}_{\hat{\alpha}}\},
\end{equation}
with energy-momentum tensor
\begin{equation}
T_{nm}=-\overline{w}^{\alpha}\partial\overline{\lambda}_{\alpha}-s^{\alpha}\partial r_{\alpha}-\hat{\overline{w}}^{\hat{\alpha}}\partial\hat{\overline{\lambda}}_{\hat{\alpha}}-\hat{s}^{\hat{\alpha}}\partial\hat{r}_{\hat{\alpha}}.
\end{equation}
The variables $\overline{\lambda}_{\alpha}$ and $\hat{\overline{\lambda}}_{\hat{\alpha}}$
are also pure spinors while $r_{\alpha}$ and $\hat{r}_{\hat{\alpha}}$
are anticommuting spinors satisfying the constraints $(\overline{\lambda}\gamma^{m}r)=0$
and $(\hat{\overline{\lambda}}\gamma^{m}\hat{r})=0$. The BRST charge
is modified accordingly,\begin{subequations}
\begin{eqnarray}
J_{BRST} & \equiv & \lambda^{\alpha}d_{\alpha}+\hat{\lambda}^{\hat{\alpha}}\hat{d}_{\hat{\alpha}}+\overline{w}^{\alpha}r_{\alpha}+\hat{\overline{w}}^{\hat{\alpha}}\hat{r}_{\hat{\alpha}},\label{eq:nmBRSTcurrent}\\
Q & \equiv & \oint\,J_{BRST},\label{eq:nmBRSTcharge}
\end{eqnarray}
\end{subequations}but this does not affect the previous cohomology
because any dependence on the non-minimal variables can be gauged
away (quartet argument).

The final step is the definition of $b_{-}$ and $b_{+}$ as
\begin{eqnarray}
b_{-} & = & \Bigg(\frac{\overline{\lambda}_{\alpha}}{(\overline{\lambda}\cdot\lambda)},G^{\alpha}\Bigg)-2!\Bigg(\frac{\overline{\lambda}_{\alpha}r_{\beta}}{(\overline{\lambda}\cdot\lambda)^{2}},H^{\alpha\beta}\Bigg)-3!\Bigg(\frac{\overline{\lambda}_{\alpha}r_{\beta}r_{\gamma}}{(\overline{\lambda}\cdot\lambda)^{3}},K^{\alpha\beta\gamma}\Bigg)\nonumber \\
 &  & +4!\Bigg(\frac{\overline{\lambda}_{\alpha}r_{\beta}r_{\gamma}r_{\lambda}}{(\overline{\lambda}\cdot\lambda)^{4}},L^{\alpha\beta\gamma\lambda}\Bigg)-s^{\alpha}\partial\overline{\lambda}_{\alpha}-\partial\Bigg(\frac{\overline{\lambda}_{\alpha}\overline{\lambda}_{\beta}}{(\overline{\lambda}\cdot\lambda)^{2}}\Bigg)\lambda^{\alpha}\partial\theta^{\beta},\label{eq:b-covariant}
\end{eqnarray}
and
\begin{eqnarray}
b_{+} & = & \Bigg(\frac{\hat{\overline{\lambda}}_{\hat{\alpha}}}{(\hat{\overline{\lambda}}\cdot\hat{\lambda})},\hat{G}^{\hat{\alpha}}\Bigg)-2!\Bigg(\frac{\hat{\overline{\lambda}}_{\alpha}\hat{r}_{\beta}}{(\hat{\overline{\lambda}}\cdot\hat{\lambda})^{2}},\hat{H}^{\hat{\alpha}\hat{\beta}}\Bigg)-3!\Bigg(\frac{\hat{\overline{\lambda}}_{\hat{\alpha}}\hat{r}_{\hat{\beta}}\hat{r}_{\hat{\gamma}}}{(\hat{\overline{\lambda}}\cdot\hat{\lambda})^{3}},\hat{K}^{\hat{\alpha}\hat{\beta}\hat{\gamma}}\Bigg)\nonumber \\
 &  & +4!\Bigg(\frac{\hat{\overline{\lambda}}_{\hat{\alpha}}\hat{r}_{\hat{\beta}}\hat{r}_{\hat{\gamma}}\hat{r}_{\hat{\lambda}}}{(\hat{\overline{\lambda}}\cdot\hat{\lambda})^{4}},\hat{L}^{\hat{\alpha}\hat{\beta}\hat{\gamma}\hat{\lambda}}\Bigg)-\hat{s}^{\hat{\alpha}}\partial\hat{\overline{\lambda}}_{\hat{\alpha}}-\partial\Bigg(\frac{\hat{\overline{\lambda}}_{\hat{\alpha}}\hat{\overline{\lambda}}_{\hat{\beta}}}{(\hat{\overline{\lambda}}\cdot\hat{\lambda})^{2}}\Bigg)\hat{\lambda}^{\hat{\alpha}}\partial\hat{\theta}^{\hat{\beta}}.\label{eq:b+covariant}
\end{eqnarray}
The last terms in $b_{-}$ and $b_{+}$ are quantum ordering contributions.

The operators $b_{-}$ and $b_{+}$ anticommute with the BRST charge
$Q$ to give the non-minimal version of $T_{-}$ and $T_{+}$:\begin{subequations}\label{eq:QbT+-}
\begin{eqnarray}
\{Q,b_{-}\} & = & T_{-}-\overline{w}^{\alpha}\partial\overline{\lambda}_{\alpha}-s^{\alpha}\partial r_{\alpha},\nonumber \\
 & \equiv & \mathcal{T}_{-}\\
\{Q,b_{+}\} & = & T_{+}-\hat{\overline{w}}^{\hat{\alpha}}\partial\hat{\overline{\lambda}}_{\hat{\alpha}}-\hat{s}^{\hat{\alpha}}\partial\hat{r}_{\hat{\alpha}}.\nonumber \\
 & \equiv & \mathcal{T}_{+}
\end{eqnarray}
\end{subequations}The demonstration of \eqref{eq:QbT+-} is a bit
lengthy because of the reordering operations. Using the operators
chain of \eqref{eq:Gdef} and \eqref{eq:HKLdef}, the $b$ ghost defined
by
\begin{equation}
b\equiv b_{-}+b_{+},\label{eq:bghosttypeII}
\end{equation}
can be shown to satisfy
\begin{equation}
\{Q,b\}=\mathcal{T}_{+}+\mathcal{T}_{-},
\end{equation}
which is equal to the energy momentum tensor of the action $S+S_{nm}$,
\begin{eqnarray}
\mathcal{T} & = & -P_{m}\partial X^{m}-p_{\alpha}\partial\theta^{\alpha}-\hat{p}_{\hat{\alpha}}\partial\hat{\theta}^{\hat{\alpha}}-w_{\alpha}\partial\lambda^{\alpha}-\hat{w}_{\hat{\alpha}}\partial\hat{\lambda}^{\hat{\alpha}}\nonumber \\
 &  & -\overline{w}^{\alpha}\partial\overline{\lambda}_{\alpha}-s^{\alpha}\partial r_{\alpha}-\hat{\overline{w}}^{\hat{\alpha}}\partial\hat{\overline{\lambda}}_{\hat{\alpha}}-\hat{s}^{\hat{\alpha}}\partial\hat{r}_{\hat{\alpha}}.
\end{eqnarray}
The existence of the $b$ ghost \eqref{eq:bghosttypeII} ensures that
the BRST cohomology is composed of worldsheet scalars only, excluding
the extra states described in subsection \ref{sub:extrastates}. Therefore,
BRST-closed massive states are unequivocally BRST-exact.

Observe that the operator $\mathcal{H}_{CV}$ defined in \eqref{eq:Hcv}
can be rewritten as
\begin{equation}
\mathcal{H}_{CV}=\{Q,(b_{+}-b_{-}+\hat{s}^{\hat{\alpha}}\partial\hat{\overline{\lambda}}_{\hat{\alpha}}-s^{\alpha}\partial\overline{\lambda}_{\alpha})\},
\end{equation}
but once the non-minimal sector is included, it makes sense to define
\begin{equation}
\mathcal{H}\equiv\mathcal{T}_{+}-\mathcal{T}_{-},
\end{equation}
which is also BRST-exact.

The properties of $b_{-}$ and $b_{+}$ are now easy to determine
because they have the same structure of the the composite $b$ ghost
of \cite{Berkovits:2005bt}. Nilpotency, for example, follows from
the same arguments of \cite{b-nilpotency} and it can be shown that
\[
b_{\pm}(z)b_{\pm}(y)\sim0.
\]
Clearly, the OPE $b_{+}(z)b_{-}(y)$ is also regular, but this follows
from the sector splitting. With respect to the BRST current, the OPE's
with $b_{\pm}$ are computed to be
\[
J_{BRST}(z)b_{\pm}(y)\sim\frac{3}{(z-y)^{3}}+\frac{J_{\pm}}{(z-y)^{2}}+\frac{\mathcal{T}_{\pm}}{(z-y)},
\]
where $J_{-}$ and $J_{+}$ are interpreted as the ghost number currents
for each sector, defined as\begin{subequations}
\begin{eqnarray}
J_{-} & \equiv & J+r_{\alpha}s^{\alpha}-2\frac{(\overline{\lambda}\cdot\partial\lambda)}{(\overline{\lambda}\cdot\lambda)}+2\frac{(r\cdot\partial\theta)}{(\overline{\lambda}\cdot\lambda)}-2\frac{(r\cdot\lambda)(\overline{\lambda}\cdot\partial\theta)}{(\overline{\lambda}\cdot\lambda)^{2}}\label{eq:J-}\\
J_{+} & \equiv & \hat{J}+\hat{r}_{\hat{\alpha}}\hat{s}^{\hat{\alpha}}-2\frac{(\hat{\overline{\lambda}}\cdot\partial\hat{\lambda})}{(\hat{\overline{\lambda}}\cdot\hat{\lambda})}+2\frac{(\hat{r}\cdot\partial\hat{\theta})}{(\hat{\overline{\lambda}}\cdot\hat{\lambda})}-2\frac{(\hat{r}\cdot\hat{\lambda})(\hat{\overline{\lambda}}\cdot\partial\hat{\theta})}{(\hat{\overline{\lambda}}\cdot\hat{\lambda})^{2}}.
\end{eqnarray}
\end{subequations}The unusual terms in $J_{\pm}$ are BRST-exact
\cite{Berkovits:2005bt} and can in fact be eliminated by a BRST transformation
of the $b$ ghost \cite{Jusinskas:2013sha}. The ghost number currents
have the following OPE's:
\begin{eqnarray}
\mathcal{T}_{\pm}(z)J_{\pm}(y) & \sim & -\frac{3}{(z-y)^{3}}+\frac{J_{\pm}}{(z-y)^{2}}+\frac{\partial J_{\pm}}{(z-y)},\\
J_{\pm}(z)J_{\pm}(y) & \sim & \frac{3}{(z-y)^{2}}.
\end{eqnarray}

Altogether, these results can be summarized as\begin{subequations}\label{eq:topoalgebra}
\begin{eqnarray}
b(z)b(y) & \sim & 0,\\
J_{BRST}(z)b(y) & \sim & \frac{6}{(z-y)^{3}}+\frac{J_{g}}{(z-y)^{2}}+\frac{\mathcal{T}}{(z-y)},\\
J_{g}(z)J_{g}(y) & \sim & \frac{6}{(z-y)^{2}},\\
\mathcal{T}(z)J_{g}\left(y\right) & \sim & -\frac{6}{(z-y)^{3}}+\frac{J_{g}}{(z-y)^{2}}+\frac{\partial J_{g}}{(z-y)},
\end{eqnarray}
\end{subequations}with
\begin{equation}
J_{g}\equiv J_{-}+J_{+}
\end{equation}
defined as the \emph{total} ghost number current.

The equations displayed in \eqref{eq:topoalgebra} resemble the $\mathcal{N}=2$
topological algebra of \cite{Berkovits:2005bt} but now with $\hat{c}=6$
and no antiholomorphic currents.

In the next section the heterotic ambitwistor string will be discussed.

\section{The heterotic ambitwistor pure spinor string\label{sec:heterotic}}

\

In \cite{Berkovits:2013xba}, Berkovits also introduced the infinite
tension limit of the heterotic pure spinor superstring. The chiral
action is given by

\begin{equation}
S=\int d^{2}z\{P_{m}\overline{\partial}X^{m}+p_{\alpha}\overline{\partial}\theta^{\alpha}+w_{\alpha}\overline{\partial}\lambda^{\alpha}+\overline{b}\overline{\partial}\overline{c}+\mathcal{L}_{C}\},\label{eq:heteroticaction}
\end{equation}
where $(\overline{b},\overline{c})$ is the known Virasoro ghost pair
for the heterotic string. $\mathcal{L}_{C}$ accounts for the Lagrangian
of the $SO(32)$ or $E(8)\times E(8)$ current algebra with central
charge $16$ and (holomorphic) generators $J^{I}$, with $I$ denoting
the adjoint representation of the gauge group. The action $S$ is
invariant under the $\mathcal{N}=1$ supersymmetry transformations
generated by the charge
\begin{equation}
q_{\alpha}=\oint\{p_{\alpha}+\frac{1}{2}P_{m}(\gamma^{m}\theta)_{\alpha}\}.\label{eq:susyheterotic}
\end{equation}

The heterotic pure spinor BRST charge was proposed to be
\begin{equation}
Q=\oint\{\lambda^{\alpha}d_{\alpha}+\overline{c}(-P_{m}\Pi^{m}-d_{\alpha}\partial\theta^{\alpha}-w_{\alpha}\partial\lambda^{\alpha}-\overline{b}\partial\overline{c}+T_{C})\},\label{eq:oldheteroticBRST}
\end{equation}
where
\begin{equation}
\Pi^{m}=\partial X^{m}+\frac{1}{2}(\theta\gamma^{m}\partial\theta),\label{eq:susymomentumheterotic}
\end{equation}
$d_{\alpha}$ is the same of \eqref{eq:dsusyinv} and $T_{C}$ is
the energy-momentum tensor associated to $\mathcal{L}_{C}$. The full
energy-momentum tensor is given by
\begin{equation}
T=-P_{m}\partial X^{m}-p_{\alpha}\partial\theta^{\alpha}-w_{\alpha}\partial\lambda^{\alpha}-\overline{b}\partial\overline{c}-\partial(\overline{b}\overline{c})+T_{C}.\label{eq:heteroticT}
\end{equation}

The massless spectrum of the heterotic string includes the non-abelian
super Yang-Mills fields and $\mathcal{N}=1$ supergravity. The former
can be encoded by the vertex operator
\begin{equation}
U_{SYM}=\lambda^{\alpha}\overline{c}A_{\alpha}^{I}(\theta)J^{I}e^{ik\cdot X},\label{eq:vertexheteroticSYM}
\end{equation}
where $A_{\alpha}^{I}(\theta)$ satisfies
\begin{equation}
D_{\alpha}\gamma_{mnpqr}^{\alpha\beta}A_{\beta}^{I}=0.
\end{equation}
The gauge transformations of $U_{SYM}$ are described by the BRST-exact
operator
\begin{eqnarray}
\delta U_{SYM} & \equiv & \{Q,\overline{c}\Lambda^{I}(\theta)J^{I}e^{ik\cdot X}\},\nonumber \\
 & = & \lambda^{\alpha}\overline{c}(D_{\alpha}\Lambda^{I})J^{I}e^{ik\cdot X}.\label{eq:gaugeheteroticSYM}
\end{eqnarray}

As for the supergravity states, the natural guess for the vertex operator
would be
\begin{equation}
U_{SG}=\lambda^{\alpha}\overline{c}A_{\alpha}^{m}(\theta)P_{m}e^{ik\cdot X}.\label{eq:heteroticsugra}
\end{equation}
BRST-closedness of $U_{SG}$ implies\begin{subequations}\label{eq:heteroticSUGRAeom}
\begin{eqnarray}
D_{\alpha}\gamma_{mnpqr}^{\alpha\beta}A_{\beta}^{s} & = & 0,\\
k_{m}A_{\alpha}^{m} & = & 0,
\end{eqnarray}
\end{subequations}which are the usual equations for the supergravity
field $A_{\alpha}^{m}$. However, the expected gauge transformation
$\delta A_{\alpha}^{m}=D_{\alpha}\Lambda^{m}+k^{m}\Lambda_{\alpha}$
does not come from a BRST-exact state:
\begin{eqnarray}
\delta U_{SG} & \equiv & \lambda^{\alpha}\overline{c}(D_{\alpha}\Lambda^{m}+k^{m}\Lambda_{\alpha})P_{m}e^{ik\cdot X},\nonumber \\
 & \neq & \{Q,\textrm{something}\}.\label{eq:wrongsugragauge}
\end{eqnarray}
Therefore, the vertex \eqref{eq:heteroticsugra} does not seem to
properly describe the heterotic supergravity spectrum \cite{Berkovits:2013xba}.

Next, motivated by the work of Chandia and Vallilo and the analysis
of the previous section, a new BRST charge for the action \eqref{eq:heteroticaction}
will be presented. The BRST cohomology will be shown to correctly
describe the massless spectrum of the heterotic string and the correspondent
$b$ ghost will be constructed.

\subsection{New proposal for the BRST charge}

\

The action \eqref{eq:heteroticaction} also has a nilpotent symmetry
that commutes with the BRST charge \eqref{eq:oldheteroticBRST}, generated
by
\begin{equation}
\mathcal{K}=\oint\,(\lambda\gamma_{m}\theta)[\partial X^{m}+\frac{1}{2}(\theta\gamma^{m}\partial\theta)].
\end{equation}
Therefore, there should be an analogous procedure to absorb this symmetry
and redefine the BRST charge consistently. 

First, the supersymmetry charge will be redefined as
\begin{equation}
q_{\alpha}=\oint\{p_{\alpha}+\frac{1}{2}(P_{m}-\partial X_{m})(\gamma^{m}\theta)_{\alpha}-\frac{1}{12}(\theta\gamma_{m}\partial\theta)(\gamma^{m}\theta)_{\alpha}\},
\end{equation}
exactly like in subsection \ref{sub:reviewCV}, which provides the
supersymmetric invariants:\begin{subequations}\label{eq:heteroticsusyinvariants}
\begin{eqnarray}
d_{\alpha} & = & p_{\alpha}-\frac{1}{2}P_{m}(\gamma^{m}\theta)_{\alpha}+\frac{1}{2}\Pi^{m}(\gamma_{m}\theta)_{\alpha},\\
P_{m}^{-} & = & P_{m}-\partial X_{m}-(\theta\gamma_{m}\partial\theta),\\
P_{m}^{+} & = & P_{m}+\partial X_{m}.
\end{eqnarray}
\end{subequations}Note that $P_{m}^{\pm}$ and $\Pi^{m}$ are not
all independent as $P_{m}^{+}=P_{m}^{-}+2\Pi_{m}$, \emph{cf.} equation
\eqref{eq:susymomentumheterotic}. In terms of the new invariants,
the action can be rewritten as
\begin{eqnarray}
S & = & \int d^{2}z\{\frac{1}{2}(P_{m}^{+}+P_{m}^{-})\overline{\Pi}^{m}+d_{\alpha}\overline{\partial}\theta^{\alpha}+\omega_{\alpha}\overline{\partial}\lambda^{\alpha}+\overline{b}\overline{\partial}\overline{c}+\mathcal{L}_{C}\}\nonumber \\
 &  & -\frac{1}{2}\int d^{2}z\{\Pi^{m}(\theta\gamma_{m}\overline{\partial}\theta)-\overline{\Pi}^{m}(\theta\gamma_{m}\partial\theta)\},\label{eq:heteroticaction2}
\end{eqnarray}
and the relevant non-regular OPE's are simply
\begin{equation}
\begin{array}{rclcrcl}
d_{\alpha}(z)d_{\beta}(y) & \sim & -\frac{P_{m}^{-}\gamma_{\alpha\beta}^{m}}{(z-y)}, &  & P_{m}^{\pm}(z)P_{n}^{\pm}(y) & \sim & \mp2\frac{\eta_{mn}}{(z-y)^{2}},\\
\\
d_{\alpha}(z)P_{m}^{-}(y) & \sim & -2\frac{(\gamma_{m}\partial\theta)_{\alpha}}{(z-y)}, &  & d_{\alpha}(z)\Pi^{m}(y) & \sim & \frac{(\gamma_{m}\partial\theta)_{\alpha}}{(z-y)},\\
\\
P_{m}^{-}(z)\Pi^{n}(y) & \sim & -\frac{\delta_{m}^{n}}{(z-y)^{2}}, &  & P_{m}^{+}(z)\Pi^{n}(y) & \sim & -\frac{\delta_{m}^{n}}{(z-y)^{2}}.
\end{array}
\end{equation}

The analogy with the type II case can be pushed further and a similar
sectorization can be shown to hold in the heterotic case. The energy-momentum
of \eqref{eq:heteroticT} can be cast as
\begin{equation}
T=T_{+}+T_{-},
\end{equation}
where\begin{subequations}
\begin{eqnarray}
T_{-} & \equiv & \frac{1}{4}\eta^{mn}P_{m}^{-}P_{n}^{-}-d_{\alpha}\partial\theta^{\alpha}-w_{\alpha}\partial\lambda^{\alpha},\\
T_{+} & \equiv & -\frac{1}{4}\eta^{mn}P_{m}^{+}P_{n}^{+}+T_{C}-\overline{b}\partial\overline{c}-\partial(\overline{b}\overline{c}),
\end{eqnarray}
\end{subequations}satisfying the same set of OPE's of \eqref{eq:partialTOPES}. 

As before, the new BRST current should naturally incorporate this
splitting and will be defined as
\begin{equation}
J_{BRST}\equiv\lambda^{\alpha}d_{\alpha}+\overline{c}(-\frac{1}{4}\eta^{mn}P_{m}^{+}P_{n}^{+}+T_{C}-\overline{b}\partial\overline{c})+\frac{3}{2}\partial^{2}\overline{c},
\end{equation}
\emph{cf.} \eqref{eq:heteroticsusyinvariants}. The last term is introduced
to make $J_{BRST}$ a conformal primary operator, but disappears in
the BRST charge $Q=\oint J_{BRST}$, such that:
\begin{equation}
Q=\oint\,\{\lambda^{\alpha}d_{\alpha}+\overline{c}T_{+}-\overline{b}\overline{c}\partial\overline{c}\}.\label{eq:newheteroticBRST}
\end{equation}
It is straightforward to show that the action is invariant under the
BRST transformations generated by \eqref{eq:newheteroticBRST}.

\subsection{BRST cohomology and the semi-composite $b$ ghost\label{sub:heteroticb}}

\

Concerning the cohomology of the BRST charge of \eqref{eq:newheteroticBRST},
only a minor modification is required. The super Yang-Mills states
are still described by the vertex $U_{SYM}$ and gauge transformation
$\delta U_{SYM}$ displayed in \eqref{eq:vertexheteroticSYM} and
\eqref{eq:gaugeheteroticSYM} respectively. On the other hand, the
$\mathcal{N}=1$ supergravity vertex will be corrected to
\begin{equation}
U_{SG}=\lambda^{\alpha}\overline{c}A_{\alpha}^{m}(\theta)P_{m}^{+}e^{ik\cdot X}.\label{eq:newheteroticsugra}
\end{equation}
BRST-closedness will again provide the equations displayed in \eqref{eq:heteroticSUGRAeom}.
The gauge transformations of $U_{SG}$ are given in terms of BRST-exact
states of the form
\begin{equation}
[Q,\Lambda]=\lambda^{\alpha}\overline{c}P_{m}^{+}(D_{\alpha}\Lambda^{m}+ik^{m}\Lambda_{\alpha})e^{ik\cdot X},
\end{equation}
where $\Lambda=2\lambda^{\alpha}\Lambda_{\alpha}-\overline{c}P_{m}^{+}\Lambda^{m}$
and $\lambda^{\alpha}\lambda^{\beta}D_{\alpha}\Lambda_{\beta}=k_{m}\Lambda^{m}=0$.

Defining $\mathbb{A}_{\alpha}^{m}(X,\theta)\equiv A_{\alpha}^{m}e^{ik\cdot X}$,
the superfield equations of motion of $U_{SG}$ can be cast as\begin{subequations}
\begin{eqnarray}
\gamma_{mnpqr}^{\alpha\beta}D_{\beta}\mathbb{A}_{\alpha}^{m} & = & 0,\\
\partial^{n}\partial_{n}\mathbb{A}_{\alpha}^{m}-\partial^{m}\partial_{n}\mathbb{A}_{\alpha}^{n} & = & 0,
\end{eqnarray}
\end{subequations}with gauge transformations given by
\begin{equation}
\delta\mathbb{A}_{\alpha}^{m}=D_{\alpha}\Lambda^{m}+\partial^{m}\Lambda_{\alpha}.\label{eq:gaugeheteroticSUGRA}
\end{equation}
As long as the gauge parameters satisfy\begin{subequations}
\begin{eqnarray}
D\gamma_{mnpqr}\Lambda & = & 0,\\
\partial^{n}\partial_{n}\Lambda^{m}-\partial^{m}\partial_{n}\Lambda^{n} & = & 0,
\end{eqnarray}
\end{subequations}\eqref{eq:gaugeheteroticSUGRA} can be seen as
a BRST transformation of $U_{SG}$, as opposed to \eqref{eq:wrongsugragauge}
in the original formulation. Note that even with this improvement
with respect to Berkovits' proposal, it is not clear whether the supergravity
theory can be recovered through these vertices. In the RNS ambitwistor
string of \cite{Mason:2013sva}, the tree level amplitudes in the
heterotic theory could not be interpreted in terms of standard space-time
gravity. This has yet to be clarified here and will be left for a
future work.

The absence of massive states in the cohomology is ensured by the
existence of a semi-composite $b$ ghost. While in \cite{Berkovits:2013xba}
the fundamental $\overline{b}$ fits the role of such operator, the
modifications introduced in the BRST charge \eqref{eq:newheteroticBRST}
imply the sectorization of the new $b$ ghost as well. Note that $\{Q,\overline{b}\}=T_{+}$,
\emph{i.e.} only part of the energy-momentum tensor \eqref{eq:heteroticT}.
Defining
\begin{equation}
b\equiv\overline{b}+b_{-},
\end{equation}
where $b_{-}$ has the same form of \eqref{eq:b-covariant} in terms
of the non-minimal variables, it is direct to show that $\{Q,b\}=\mathcal{T}$,
with
\begin{eqnarray}
\mathcal{T} & = & -P_{m}\partial X^{m}-p_{\alpha}\partial\theta^{\alpha}-w_{\alpha}\partial\lambda^{\alpha}+T_{C}\nonumber \\
 &  & -\overline{b}\partial\overline{c}-\partial(\overline{b}\overline{c})-\overline{w}^{\alpha}\partial\overline{\lambda}_{\alpha}-s^{\alpha}\partial r_{\alpha}.
\end{eqnarray}
 The heterotic $b$ ghost consists of a fundamental part $\overline{b}$
and the usual (pure spinor) composite one, $b_{-}$.

For completeness, the heterotic case can be shown to have a similar
OPE set as the one displayed in \eqref{eq:topoalgebra},\begin{subequations}
\begin{eqnarray}
b(z)b(y) & \sim & 0,\\
J_{BRST}(z)b(y) & \sim & \frac{6}{(z-y)^{3}}+\frac{J_{g}}{(z-y)^{2}}+\frac{\mathcal{T}}{(z-y)},\\
J_{g}(z)J_{g}(y) & \sim & \frac{4}{(z-y)^{2}},\\
\mathcal{T}(z)J_{g}\left(y\right) & \sim & -\frac{6}{(z-y)^{3}}+\frac{J_{g}}{(z-y)^{2}}+\frac{\partial J_{g}}{(z-y)},
\end{eqnarray}
\end{subequations}with
\begin{equation}
J_{g}\equiv J_{-}+cb,
\end{equation}
where $J_{-}$ is defined in \eqref{eq:J-}.

\section{Final remarks\label{sec:remarks}}

\

The results presented in \cite{Chandia-Vallilo} and developed here
have to be further explored, but it is interesting to see that the
sectorization of the holomorphic $\alpha'\to0$ limit of the pure
spinor superstring describes the expected massless spectrum in a very
simple way and enables a natural definition for the composite $b$
ghost.

The geometrical interpretation of ($b_{+}$,$b_{-}$) for type II
and ($\overline{b}$,$b_{-}$) for heterotic theories is not clear
yet. The ideas of \cite{Adamo:2013tsa} might shed some light in the
pure spinor construction, since there the 1-loop scattering equations
of the ambitwistor formulation for the RNS string were studied in
detail. To make it more precise, notice that in \cite{Mason:2013sva},
the operators $b$ and $\tilde{b}$ satisfy\begin{subequations}
\begin{eqnarray}
\{Q,b\} & = & T,\\
\{Q,\tilde{b}\} & = & \frac{1}{2}P_{m}P^{m},
\end{eqnarray}
\end{subequations}while in the pure spinor case discussed here one
has\begin{subequations}
\begin{eqnarray}
\{Q,b\} & = & T,\\
\{Q,\tilde{b}\} & = & T_{+}-T_{-},\nonumber \\
 & = & \frac{1}{2}P_{m}P^{m}+\ldots.\label{eq:qbtilde}
\end{eqnarray}
\end{subequations}The operator $\tilde{b}$ is defined as\begin{subequations}\label{eq:btildePS}
\begin{eqnarray}
\tilde{b}_{II} & \equiv & b_{+}-b_{-},\\
\tilde{b}_{het} & \equiv & \overline{b}-b_{-},
\end{eqnarray}
\end{subequations}according to the results of subsections \ref{sub:typeIIbghost}
and \ref{sub:heteroticb}. Since the parallel is clear, a natural
step now would be to investigate the consistency (\emph{e.g.} modular
invariance) of the 1-loop amplitude prescription in the same line
of \cite{Adamo:2013tsa} with the adequate identifications. In \cite{Adamo:2015hoa},
an amplitude prescription was presented following Berkovits' proposal
\cite{Berkovits:2013xba}. There, because of the absence of a true
$b$ ghost satisfying $\{Q,b\}=T$, BRST-invariance of the amplitude
does not have the usual surface terms in the \emph{moduli} space integration
but is achieved through the $\overline{\delta}(P^{2})$ insertions
proposed in \cite{Adamo:2013tsa}, much like BRST invariance of Berkovits'
integrated vertex depends on the $\overline{\delta}(k\cdot P)$ operator
(see equation \eqref{eq:integratedvertexB}). It would be interesting
to have an alternative prescription using the sectorized construction
and to compare both approaches.

From another perspective, the operators $\tilde{b}_{II}$ and $\tilde{b}_{het}$
of \eqref{eq:btildePS} seem to provide the analogous of the physical
state condition in the closed string,
\begin{equation}
(b_{L})_{0}-(b_{R})_{0}\left|\psi\right\rangle =0.
\end{equation}
The index $0$ denotes the zero mode of the left and right-moving
$b$ ghost of the closed string. Physical states here will then be
defined as elements of the BRST cohomology satisfying\begin{subequations}
\begin{eqnarray}
(\tilde{b}_{II})_{0}\left|\psi\right\rangle  & \approx & 0,\\
(\tilde{b}_{het})_{0}\left|\psi\right\rangle  & \approx & 0.
\end{eqnarray}
\end{subequations}The symbol $\approx$ means equal up to BRST-exact
terms.

The integrated form of the vertex operators is still lacking, but
the sectorized $b$ ghost operators ($b_{\pm}$ and $\overline{b}$)
might play an important role. In \cite{Adamo:2013tsa} there is a
direct relation between the integrated and the unintegrated vertices
through $b$ and $\tilde{b}$ insertions. It is very likely that a
similar construction can be found here to build the integrated vertices
associated to \eqref{eq:sugraIIvertex}, \eqref{eq:vertexheteroticSYM}
and \eqref{eq:newheteroticsugra}. This idea has to be further developed
and certainly deserves more attention for a precise formulation of
the ambitwistor string in the pure spinor formalism.

Last, concerning the heterotic case, there is a very straightforward
test for the new BRST charge proposed in \eqref{eq:newheteroticBRST}.
The heterotic action
\begin{eqnarray}
S_{het} & = & \int d^{2}z\{\frac{1}{2}(P_{m}^{+}+P_{m}^{-})\overline{\Pi}^{m}+d_{\alpha}\overline{\partial}\theta^{\alpha}+\omega_{\alpha}\overline{\partial}\lambda^{\alpha}+b\overline{\partial}c+\mathcal{L}_{C}\}\nonumber \\
 &  & -\frac{1}{2}\int d^{2}z\{\Pi^{m}(\theta\gamma_{m}\overline{\partial}\theta)-\overline{\Pi}^{m}(\theta\gamma_{m}\partial\theta)\},
\end{eqnarray}
given in terms of the redefined supersymmetric invariants of \eqref{eq:heteroticsusyinvariants},
naturally presents the coupling with the Kalb-Ramond field in the
zero-momentum limit (second line), analogous to Chandia and Vallilo's
proposal for the type II case \cite{Chandia-Vallilo}. Therefore,
the curved background embedding of $S_{het}$ should provide the heterotic
supergravity constraints of \cite{Berkovits:2001ue} through a sensible
curved space generalization of $T_{\pm}$ \cite{Azevedo-Jusinskas}.
As mentioned in section \ref{sec:heterotic}, the RNS ambitwistor
string does not provide the expected supergravity amplitudes. A similar
analysis will have to be performed here. It seems, however, that one
might expect similar results and possible inconsistencies could show
up in determining the supergravity constraints.

\textbf{Acknowledgements:} I would like to thank Thales Azevedo for
useful comments and discussions. Also, Nathan Berkovits and Andrei
Mikhailov for reading the draft. This research has been supported
by the Grant Agency of the Czech Republic, under the grant P201/12/G028.

\end{document}